\documentclass[review,11pt]{elsarticle}

\biboptions{sort&compress}  

\usepackage[mathscr]{eucal}  
\usepackage{graphicx,multirow} 
\usepackage[normal]{subfigure}
\usepackage{amssymb,amsmath} 
\usepackage{accents}
\usepackage{mathtools}  
\usepackage{booktabs}
\usepackage{cancel}

\usepackage{tikz}  

\newcommand{\be}{\begin{eqnarray}}
\newcommand{\ee}{\end{eqnarray}}
\newcommand{\nn}{\nonumber}
\newcommand{\mbf}[1]{\mbox{$\mathbf #1$}}    
\newcommand{\mref}[1]{(\ref{#1})}

\DeclareMathOperator*{\argmin}{arg\,min}

\usepackage[linesnumbered,ruled]{algorithm2e}
\newif\iflongver 
\longverfalse  

\journal{Signal Processing}

\begin{document}


\begin{frontmatter}
\title{Delta-Iterations: A General-Purpose Linear Regression Solver for Huber's Penalty, $\epsilon$-Insensitive Loss, and Other Moreau Envelope-Based Loss Functions}

\author[METU]{\c{C}a\u{g}atay Candan}
\ead{ccandan@gmail.com}

\address[METU]{Department of Electrical and Electronics Engineering, Bogazici University, Bebek, Istanbul, 34342, Turkey}

\begin{abstract}
A general-purpose linear regression solver for a class of smooth loss functions is presented.
The method applies to loss functions of the Moreau envelope type, which are continuously differentiable and have Lipschitz-continuous first derivatives. This class includes piecewise-defined cost functions consisting of constant, linear and quadratic segments, such as Huber's loss, insensitive least squares, one-sided least squares. The convergence guarantee provided by the associated proximal operator, the simplicity of the method and its versatility in handling various cost functions within a unified framework are its main advantages. MATLAB code for reproducing the numerical results presented in this article is provided.
\end{abstract}

\begin{keyword}
Linear Regression, Least Squares, Huber's penalty, Insensitive least squares, Moreau Envelope, Proximal Operator.
\end{keyword}
\end{frontmatter}

\section{Introduction}
The quadratic cost function and related least-squares techniques are foundational to virtually all areas of signal processing. Beyond quadratic cost functions, several other penalties serve different purposes. These include Huber's penalty (a piecewise-defined function consisting of quadratic and linear segments) for robust regression \cite{HuberSolver2026,HuberADMM}, the $\epsilon$-insensitive cost (a piecewise-defined function consisting of constant and quadratic segments) \cite{Karal2023epsins,Huang2005epsins} for handling bounded noise, and the $\ell_1$ norm \cite{L1normRegression2018,LASSOBook} for promoting sparsity. A major disadvantage of non-quadratic cost functions is that they often require specialized solvers tailored to the particular problem, such as the celebrated ISTA/FISTA methods for LASSO regression \cite{FISTABeck,LASSOBook}.

In this study, we propose a general iterative solver, called delta-iterations, for a class of cost functions that includes Huber's loss, the insensitive cost, and many others. More specifically, when the cost function is a Moreau envelope (i.e., a convex, continuously differentiable function with a bounded second derivative or, more generally, a Lipschitz-continuous gradient; see \cite[Ch. 6.7]{BeckFOoptBook}), delta-iterations take the conventional least-squares solution as the initial estimate and perform a fixed-point iteration that converges to the optimal solution of the non-quadratic cost function. The algorithm computes the {\em difference} between the conventional least-squares solution (the initial point) and the optimal solution, hence the name delta-iterations. The ability to handle multiple cost functions within the same framework, together with the convergence guarantee and low computational complexity, makes delta-iterations a practical alternative to existing solvers in many applications. 

The Moreau envelope is a smooth (continuously differentiable) approximation of a convex function \cite[Ch. 6.7]{BeckFOoptBook}. The minimizers of the Moreau envelope and of the possibly nonsmooth convex function from which it is generated, hereafter referred to as the seed function, are identical \cite[p. 131]{BoydProxBook}. For example, Huber's penalty is a smooth approximation of the $\ell_1$ norm \cite[Ex. 10.54]{BeckFOoptBook}. Despite the popularity of Huber's penalty in regression, few studies have exploited the general properties of Moreau envelopes in signal processing applications, with the exception of Selesnick's work on image denoising. Selesnick proposed using the difference between the $\ell_1$ norm and its Moreau envelope (i.e., the Moreau envelope approximation error) as a nonconvex regularizer to improve sparse representation \cite{SelMEnvTSP2017,Sel2017SPL}. Similar to the approach followed in the present work, Selesnick also proposes using the proximal gradient method (also known as the forward-backward splitting algorithm) for efficient optimization.

We present four different algorithms in this study: Algorithm~\ref{Alg1}, which addresses overdetermined linear equation systems with a Moreau envelope cost function on the residual error; Algorithm~\ref{AlgSQUAREM}, which is a SQUAREM-accelerated version of Algorithm~\ref{Alg1} with Tikhonov regularization; Algorithm~\ref{AlgMenvReg}, which addresses conventional least-squares linear regression with a Moreau envelope regularizer; and Algorithm~\ref{AlgMinMenvNorm}, which computes the minimum Moreau envelope solution of an underdetermined linear system. Signal processing applications of algorithms are presented in the numerical results section.


\begin{table*}[]
\caption{Seed function $h(u)$, proximal operator and Moreau envelope}
\centering
\begin{tabular}{|ccc|} \hline
\vspace*{-0.2cm}& & \\
$h(u)$ & $\textrm{prox}_{\mu h}(x)$ &$M_h^\mu(x)$ \\
\vspace*{-0.2cm}& & \\ \hline
\vspace*{-0.2cm}& & \\
$\frac{1}{2} u^2$ & $\frac{1}{1+\mu}x$ & $\frac{1}{2(1+\mu)}x^2$ \\
\vspace*{-0.2cm}& & \\
$|u|$ &
$S_\mu(x) = \left\{
\begin{array}{cc}
x+\mu & x \leq -\mu \\
0     & |x| \leq \mu \\
x-\mu & x \geq \mu
\end{array} \right.$ &
$H_\mu(x) = \left\{
\begin{array}{cc}
\frac{1}{2\mu}x^2      & |x| \leq \mu \\
|x| - \frac{\mu}{2} & |x| \geq \mu
\end{array} \right.$ \\
\vspace*{-0.2cm}& & \\
$I_{|u|\leq\lambda}(u) =
\left\{
\begin{array}{cc}
0      & |u| \leq \lambda \\
\infty & |u| > \lambda
\end{array} \right.$
&
$C_\lambda(x) = \left\{
\begin{array}{cc}
-\lambda     & x \leq -\lambda \\
x  & |x| \leq \lambda \\
\lambda & x \geq \lambda
\end{array} \right.$ &
$L_\lambda(x) = \left\{
\begin{array}{cc}
0 & |x| \leq \lambda \\
\frac{1}{2\mu}(|x| - \lambda)^2 & |x| \geq \lambda
\end{array} \right.$ \\
$I_{u\geq 0}(u) =
\left\{
\begin{array}{cc}
0      & u \geq 0 \\
\infty & u < 0
\end{array} \right.$
&
$[x]_{\text{+}} = \max(x,0)$
&
$ J_{\mu}(x) = \left\{
\begin{array}{cc}
0 & x \geq 0 \\
\frac{1}{2\mu}x^2 & x \leq 0
\end{array} \right.$
\\ \hline
\end{tabular}
\label{TableM}
\end{table*}

\vspace*{-0.4cm}
\section{Background}
The Moreau envelope of a convex function $h(u) :  \mathbb{E} \rightarrow \mathbb{R}$ is defined as \cite[Ch. 10.8.3]{BeckFOoptBook}
\be
M_h^\mu(x) = \min_{u \in \mathbb{E}} \left\{ h(u) + \frac{1}{2\mu} | x - u |^2 \right\}
\label{moreaudef}
\ee
The Moreau envelope is a convex, smooth (continuously differentiable) function that approximates $h(u)$. We refer to $h(u)$ as the seed function that generates the envelope. For example, Huber's penalty $H_\mu(x)$ (see Table~\ref{TableM}) is the Moreau envelope of the seed function $h(x)=|x|$ \cite{HastieEOSLBook}. As the smoothing parameter $\mu>0$ decreases, the envelope approaches $h(x)$, as can be seen from  Table~\ref{TableM}.

In addition, the approximation $M_h^\mu(x) :  \mathbb{R} \rightarrow \mathbb{R}$ lifts the restriction on the domain of $h(u) :  \mathbb{E} \rightarrow \mathbb{R}$.   For example, the $\lambda$-insensitive cost function $L_\lambda(x)$ shown in Table~\ref{TableM} has a seed function whose domain is restricted to $[-\lambda,\lambda]$, whereas its Moreau envelope is well defined and finite for all $x\in\mathbb{R}$. The smoothness and full-domain property of the Moreau envelope are important for many numerical methods, including the proximal gradient method \cite{BoydProxBook} used in this work.

The proximal function associated with $h(u)$ is defined as
\be
\textrm{prox}_{\mu h}(x) = \argmin_{u \in \mathbb{E}} \left\{ h(u) + \frac{1}{2\mu} | x - u |^2 \right\}.
\label{proxdef}
\ee
By comparing \mref{moreaudef} and \mref{proxdef}, we see that the Moreau envelope and the proximal mapping correspond to the optimal value and the optimizer of the same optimization problem, respectively. Hence, we have $M_h^\mu(x) = h( \textrm{prox}_{\mu h}(x) ) + \frac{1}{2\mu} | x - \textrm{prox}_{\mu h}(x) |^2$.

For a vector-valued $\mbf{x} = [x_1\, x_2\, \ldots \, x_K]^T$, we define the Moreau envelope $M^{\mu}_{h}(\mathbf{x}) : \mathbb{R}^K \rightarrow \mathbb{R}$ as an additively separable function: $M^{\mu}_{h}(\mathbf{x}) \triangleq \sum_{k=1}^K M^{\mu}_{h}(x_k)$. For example, the Huber's penalty with a vector argument is given by $\sum_{k=1}^K H_{\mu}(x_k)$, that is the penalty is applied to each element of $\mbf{x}$ and the resulting values are summed. Similarly, for a vector-valued $\mbf{x}$, the proximal mapping $\textrm{prox}_{\mu h}(\mbf{x}) : \mathbb{R}^K \rightarrow \mathbb{R}^K$ is applied individually to each element $\mbf{x}$.

\vspace*{-0.4cm}
\section{Main Result}
We consider the linear regression problem with a Moreau envelope loss function $M^{\mu}_{h}(\mathbf{x})$
\be
\mathbf{x}_* = \argmin_{\mathbf{x} \in \mathbb{R}^K} M^{\mu}_{h}(\mbf{b} - \mathbf{A}\mathbf{x}).
\label{eqMR1}
\ee
Here, $\mathbf{A}$ is an $N \times K$ matrix and $\mbf{b}$ is an $N \times 1$ vector, with $N>K$.  Substituting  the variational definition of $M^{\mu}_{h}(\cdot)$ from \mref{moreaudef} into \mref{eqMR1}, we obtain
\be
 \min_{\mathbf{x} \in \mathbb{R}^K}
 M^{\mu}_{h}(\mbf{b} - \mathbf{A}\mathbf{x})
 \!\!=\!\!
\min_{\mathbf{x} \in \mathbb{R}^K}
\min_{\mathbf{u} \in \mathbb{E}}
\underbrace{
\{ h(u) + \frac{1}{2\mu} \| \mbf{b} - \mathbf{A}\mathbf{x} - \mbf{u} \|^2 \}}_{J(\mathbf{x},\mathbf{u})}
\nn
\ee
where ${J(\mathbf{x},\mathbf{u})}$ denotes the cost function for the joint minimization over $\mbf{x}$ and $\mbf{u}$.

Using the joint optimization formulation, we can reverse the order of minimization and first minimize with respect to $\mbf{x}$ for a fixed $\mbf{u}$. The resulting solution is
\be
\hat{\mathbf{x}}(\mbf{u}) & = &  (\mathbf{A}^T\mathbf{A})^{-1}\mathbf{A}^T(\mathbf{b}-\mathbf{u}) = \mbf{A}^\dagger (\mathbf{b}-\mathbf{u})
=
\mathbf{x}_{\textrm{OLS}} - \mbf{A}^\dagger \mathbf{u}
\label{OLS}
\ee
where $\mbf{A}^\dagger = (\mathbf{A}^T\mathbf{A})^{-1}\mathbf{A}^T$ is the pseudo-inverse matrix and $\mathbf{x}_{\textrm{OLS}} = \mbf{A}^\dagger \mathbf{b}$ is the ordinary least squares solution. Substituting $\hat{\mathbf{x}}(\mbf{u})$ into the cost function, we get
\be
u_* & = &  \argmin_{\mathbf{u} \in \mathbb{E}}
J(\hat{\mathbf{x}}(\mbf{u}), \mathbf{u})
 =  \argmin_{\mathbf{u} \in \mathbb{E}}
\left\{ \frac{1}{2}
\| \mathbf{P}_\mathbf{A}^\perp (\mbf{b} -\mbf{u})\|^2
+ \mu h(\mathbf{u})
\right\}
\label{optprobu}
\ee
where $\mbf{P}_\mathbf{A} = \mathbf{A}
(\mathbf{A}^T\mathbf{A})^{-1}\mathbf{A}^T$ and
$\mathbf{P}_\mathbf{A}^\perp = \mbf{I} - \mbf{P}_\mathbf{A}$ is the orthogonal projection matrix to the range space of $\mathbf{A}$ and its orthogonal complement, respectively.

To handle the minimization over $\mbf{u}$ variable in \mref{optprobu}, we decompose the objective as
$ \min_{\mathbf{u} \in \mathbb{E}} (g(\mathbf{u}) + \mu h(\mathbf{u})$ where $g(\mathbf{u}) = \frac{1}{2}\| \mathbf{P}_\mathbf{A}^\perp (\mbf{b} - \mbf{u})\|^2$ is the smooth component of the objective, while $\mu h(\mathbf{u})$ is its possibly non-smooth component. This decomposition allows us to apply the proximal gradient method \cite{BoydProxBook}. If we denote $\mathbf{u}^{(k)}$ as the current proximal gradient iterate, the next iterate is computed using the following recursion:
\be
\mathbf{u}^{(k+1)} =
\argmin_{\mathbf{u} \in \mathbb{E}}
\frac{1}{2\eta}\|\mathbf{u} -
(\mathbf{u}^{(k)} - \eta \nabla g(\mathbf{u}^{(k)})\|^2
+ \mu h(\mathbf{u})
\label{proxg}
\ee
Here $\mathbf{u}^{(k)} - \eta \nabla g(\mathbf{u}^{(k)})$ is the gradient descent update for the smooth component of the objective, with the step-size $\eta$. For $g(\mathbf{u}) = \frac{1}{2} \| \mathbf{P}_\mathbf{A}^\perp (\mbf{u}-\mbf{b})\|^2$, we have  $\nabla g(\mathbf{u}^{(k)}) = \mathbf{P}_\mathbf{A}^\perp ( \mbf{u^{(k)}}-\mbf{b})$. Since $\mathbf{P}_\mathbf{A}^\perp$ is an orthogonal projection matrix, its eigenvalues are either $0$ or $1$. Hence, the Lipschitz constant of $\nabla g$ is $1$, and the step size can be chosen as $\eta=1$, which is the reciprocal of the largest eigenvalue of the Hessian, $\nabla^2 g(\mbf{u})=\mathbf{P}\mathbf{A}^\perp$ \cite{BoydProxBook}. Substituting the gradient update into \mref{proxg}, we obtain
\be
\mathbf{u}^{(k+1)} & = &
\argmin_{\mathbf{u} \in \mathbb{E}}
\frac{1}{2\mu}\|\mathbf{u} -
(\mathbf{P}_\mathbf{A}\mathbf{u}^{(k)} + \mathbf{P}_\mathbf{A}^\perp \mathbf{b})\|^2
+ h(\mathbf{u}) \nn \\
& = &
\textrm{prox}_{\mu h}(\mathbf{P}_\mathbf{A}\mathbf{u}^{(k)} + \mathbf{P}_\mathbf{A}^\perp \mathbf{b}) \nn \\
& = &
\textrm{prox}_{\mu h}(\mathbf{P}_\mathbf{A}\mathbf{u}^{(k)} + \mathbf{r}_\textrm{OLS})
\label{proxg2}
\ee
where the proximal operator definition in \mref{proxdef} is used in the second line and $\mathbf{r}_\textrm{OLS} = \mathbf{P}_\mathbf{A}^\perp \mathbf{b} = \mbf{b} - \mbf{A}\mbf{x}_{\textrm{OLS}}$ is the residual of the ordinary least squares solution. This completes the derivation of the recursion for $\mbf{u}$. As a final note, we emphasize that the
chosen step size $\eta=1$ for the proximal gradient iterations follows from the particular structure of the smooth component of the objective function rather than being arbitrary.

We note that \mref{proxg2} is a fixed-point iteration. The non-expansive property of the proximal operator and the choice of step size $\eta=1$ guarantees the convergence of the fixed-point iterations \cite[Ch.6.5]{BeckFOoptBook}. Hence, the proximal gradient iterations in \mref{proxg2} are continued until the stopping criterion is satisfied.

Let $\mbf{u}_*$  denote the fixed-point of the proximal gradient iteration, which corresponds to the optimal solution for $\mathbf{u}$). The optimal solution for $\mbf{x}$ is then given by (see \mref{OLS}),
\be
\mbf{x}_* =   \hat{\mathbf{x}}(\mbf{u}_*)
& = &
\mathbf{A}^\dagger \mbf{b} - \mathbf{A}^\dagger \mbf{u}_* \nn 
=
\mathbf{x}_{\textrm{OLS}} - \boldsymbol{\delta}_*
\ee
where we introduce $\boldsymbol{\delta}_* = \mathbf{A}^\dagger \mbf{u}_*$.

To reduce the computational load associated with the $\mbf{u}$-iterations in \mref{proxg2}, we introduce the variable $\boldsymbol{\delta}^{(k)} = \mathbf{A}^\dagger \mbf{u}^{(k)} \in \mathbb{R}^K$ and reformulate the fixed-point iteration in terms of $\boldsymbol{\delta}^{(k)}\in\mathbb{R}^K$ rather than $\mbf{u}^{(k)}\in\mathbb{R}^N$. Since we have an overdetermined linear equation system with $\mbf{A} \in \mathbb{R}^{N \times K}$ and $K<N$; this reformulation enables us to reduce the dimension of the iterated variable from $N$ to $K$.

We observe that the definition $\boldsymbol{\delta}^{(k)} = \mathbf{A}^\dagger \mbf{u}^{(k)}$ immediately leads to $\mbf{A} \boldsymbol{\delta}^{(k)} = \mbf{A} \mathbf{A}^\dagger \mbf{u}^{(k)} =  \mathbf{P}_\mathbf{A} \mbf{u}^{(k)}$.  Left-multiplying both sides of the proximal gradient iteration in \mref{proxg2} with $\mathbf{A}^\dagger$, we get the delta-iterations:
\be
\boldsymbol{\delta}^{(k+1)}  =
\mbf{A}^\dagger \textrm{prox}_{\mu h}(\mathbf{A}\boldsymbol{\delta}^{(k)} + \mbf{r}_{\textrm{OLS}})
\label{deltaiter}.
\ee
Upon convergence of the delta-iterations \mref{deltaiter} to $\boldsymbol{\delta}_*$, the final solution is obtained as $\mbf{x}_* = \mbf{x}_{\textrm{OLS}} - \boldsymbol{\delta}_*$ and $\mbf{u}_* =
\textrm{prox}_{\mu h}(\mathbf{A}\boldsymbol{\delta}_* + \mbf{r}_{\textrm{OLS}})$. Note that $\boldsymbol{\delta}_*$ represents the difference between the ordinary least squares solution and the optimal solution of the linear regression problem with the non-quadratic cost function. The method is summarized in Algorithm~\ref{Alg1}.\

\begin{algorithm}[t]
\label{Alg1}
\caption{\small{Linear Regression Solver (see \cite{deltaiterationsMATLABcode})}  \\
\hspace*{2.4cm} $\mbf{x}_* = \argmin M_h^\mu(\mbf{b} - \mbf{A}\mbf{x})$}
\SetKwInOut{Input}{Input}
\SetKwInOut{Output}{Output}
\SetKwInOut{Return}{Return}
{\small\Input{$\mathbf{A} \in \mathbb{R}^{N \times K} (N > K), \mbf{b}$, $\textrm{prox}_{\mu h} (x)$}

\Output{$\mbf{x}_*$}
     $\mathbf{K} = (\mbf{A}^T\mbf{A})^{-1}\mbf{A}^T$ \tcp*{Pseudo-inverse}

     $\mathbf{x}_\textrm{OLS} = \mathbf{K}\mbf{b}$
     \tcp*{Ordinary Least Squares}

     $\mathbf{r}_\textrm{OLS} = \mbf{b} - \mbf{A}\mathbf{x}_\textrm{OLS}$\tcp*{residual of OLS}

     $\boldsymbol{\delta}^{\textrm{new}} = \textrm{zeros(size(}\mathbf{x}_\textrm{OLS}))$ \tcp*{initialization}

    \Repeat{$\|\boldsymbol{\delta}^{\text{new}} - \boldsymbol{\delta}^{\text{old}}\|_\infty < \epsilon$}
    {
      $\boldsymbol{\delta}^{\textrm{old}} = \boldsymbol{\delta}^{\textrm{new}} $ \tcp*{delta-iterations}

      $\boldsymbol{\delta}^{\textrm{new}} = \mathbf{K}\textrm{prox}_{\mu h}(\mathbf{A}\boldsymbol{\delta}^{\textrm{old}} + \mathbf{r}_\textrm{OLS})$;
    }

	\Return{$\mathbf{x}_* = \mathbf{x}_\textrm{OLS} - \boldsymbol{\delta}^{\textrm{new}}$}

}
\end{algorithm}

\vspace*{-0.4cm}
\section{Extensions to related problems}
In this section, we present several extensions of the main result. The proofs of these extensions follow a similar approach to that of the main result. In each case, the proof consists of three steps: (i) formulating the joint optimization problem over $\mbf{x}$ and $\mbf{u}$; (ii) optimizing with respect to $\mbf{x}$ for a fixed $\mbf{u}$; and (iii) applying the proximal gradient method to optimize with respect to $\mbf{u}$. The main challenge lies in deriving a solution to the quadratic optimization problem in step (ii) that admits a suitable form for the subsequent application of the proximal gradient method in step (iii). For completeness, the detailed proofs are provided in appendices. 

\textbf{Extension 1: SQUAREM Acceleration}
To accelerate the $O(1/k)$ convergence rate of Algorithm~\ref{Alg1} \cite[Ch. 10.4.3]{BeckFOoptBook}, we employ the SQUAREM acceleration scheme \cite{SQUAREM-Stoica}. SQUAREM extrapolates successive fixed-point iterates, as shown in line~10 of Algorithm~\ref{AlgSQUAREM}, using a specially chosen step size determined by the first- and second-order differences between the iterates. In our numerical experiments, SQUAREM reduces the number of iterations by approximately an order of magnitude (from hundreds to tens), as demonstrated in the next section.

\textbf{Extension 2: Moreau Envelope Regression with $\ell_2$-regularization (Algorithm 2) } The proposed delta-iterations can be extended to the $\ell_2$-regularized linear regression problem
\be
\min_{\mathbf{x} \in \mathbb{R}^K} \left\{
M^{\mu}_{h}(\mbf{b} - \mathbf{A}\mathbf{x}) + \gamma \|\mathbf{x}\|^2  \right\}
\ee
by replacing the line-1 of Algorithm~\ref{Alg1} with
$\mathbf{K} = (\mbf{A}^T\mbf{A} + 2\mu\gamma \mathbf{I})^{-1}\mbf{A}^T$. Please see the appendix for details.

\begin{algorithm}[t]
\label{AlgSQUAREM}
\caption{\small{SQUAREM-accelerated Solver (see \cite{deltaiterationsMATLABcode})} \\ \hspace*{2.2cm} $\mathbf{x}_* = \argmin M_h^\mu(\mathbf{b} - \mathbf{A}\mathbf{x}) + \gamma\|\mathbf{x}\|_2^2$}
\SetKwInOut{Input}{Input}
\SetKwInOut{Output}{Output}
\SetKwInOut{Return}{Return}
{\small\Input{$\mathbf{A}, \mbf{b}, \gamma, \mu$, $\textrm{prox}_{\mu h}(x)$

\Output{$\mbf{x}_*$}

$\mathbf{K} = (\mathbf{A}^T\mathbf{A} + 2\mu\gamma\mathbf{I})^{-1}\mathbf{A}^T$;\tcp{\!\!$\gamma = 0$\!\! : unregularized}

     $\mathbf{x}_{\text{RLS}} = \mathbf{K}\mathbf{b};\,
     \mathbf{r}_{\text{RLS}} = \mathbf{b} - \mathbf{A}\mathbf{x}_{\text{RLS}};\,
     \boldsymbol{\delta}^{\text{new}} =
     \textrm{zeros(size(}\mathbf{x}_\textrm{RLS}));$

    \Repeat{$\|\boldsymbol{\delta}^{\text{new}} - \boldsymbol{\delta}^{\text{old}}\|_\infty < \epsilon$}
    {
      $\boldsymbol{\delta}^{\textrm{old}} = \boldsymbol{\delta}^{\textrm{new}};$

            $\tilde{\boldsymbol{\delta}}_1 = \mathbf{K}\,\textrm{prox}_{\mu h}(\mathbf{A}\boldsymbol{\delta}^{\text{old}} + \mathbf{r}_{\text{RLS}})$; \tcp{1st probe}

      $\tilde{\boldsymbol{\delta}}_2 = \mathbf{K}\,\textrm{prox}_{\mu h}(\mathbf{A}\tilde{\boldsymbol{\delta}}_1 + \mathbf{r}_{\text{RLS}})$; \tcp{2nd probe}

      $\mathbf{q}_1 = \tilde{\boldsymbol{\delta}}_1 - \boldsymbol{\delta}^{\text{old}}$; \tcp{1st-order difference}

      $\mathbf{q}_2 = \tilde{\boldsymbol{\delta}}_2 - \tilde{\boldsymbol{\delta}}_1 - \mathbf{q}_1$; \tcp{2nd-order difference}

      $\alpha = \min\left(-1, \, -\|\mathbf{q}_1\|_2 / \|\mathbf{q}_2\|_2\right)$; \tcp{step-size}

      $\boldsymbol{\delta}_{\text{ext}} = \boldsymbol{\delta}^{\text{old}} - 2\alpha\mathbf{q}_1 + \alpha^2\mathbf{q}_2$; \tcp{extrapolation}

      $\boldsymbol{\delta}^{\text{new}} = \mathbf{K}\,\textrm{prox}_{\mu h}(\mathbf{A}\boldsymbol{\delta}_{\text{ext}} + \mathbf{r}_{\text{RLS}})$; \tcp{new estimate}
    }
	\Return{$\mathbf{x}_* = \mathbf{x}_{\text{RLS}} - \boldsymbol{\delta}^{\text{new}}$}	
}}
\end{algorithm}

\textbf{Extension 3: $\ell_2$-regression with a Moreau Envelope Regularizer (Algorithm 3)}
The regularization term in linear regression problems can be interpreted as incorporating a priori information about the unknown vector $\mbf{x}$ in a Bayesian framework. Conventional regularized least-squares regression can be extended by using a Moreau envelope as the regularizer, as shown below:
\be
\min_{\mathbf{x}} \frac{1}{2}\|\mbf{A}\mbf{x} - \mbf{b}\|^2 + \lambda M_{h}^{\mu}(\mbf{x}).
\label{regmoreau}
\ee
The delta-iteration solution to this problem is given in Algorithm~\ref{AlgMenvReg} (see appendices for details). As an application example, Algorithm~\ref{AlgMenvReg} can be used to solve the non-negative least-squares (NNLS) problem, as demonstrated in the numerical results section.

\textbf{Extension 4: Minimum Moreau Envelope Solution of an Underdetermined Linear Equation System (Algorithm 4)}
We extend the conventional minimum $\ell_2$-norm solution problem for an underdetermined linear equation system to the  Moreau envelope setting:
\be
\min_{\mathbf{x}}
M_h^\mu(\mathbf{x}) \quad \textrm{s.t.} \quad  \mathbf{A}\mathbf{x} = \mathbf{b}.
\ee
Here $\mbf{A}$ is an $N \times K$ matrix with $K>N$. Hence, the linear system $\mbf{A}\mbf{x} = \mbf{b}$ is
underdetermined and provided that it is consistent, has infinitely many solutions. The delta-iteration solution to this problem is given in Algorithm~\ref{AlgMinMenvNorm}. Further details are provided in appendices. 

\begin{algorithm}
\label{AlgMenvReg}
\caption{\small{Regression with Moreau Env. Regularizer \cite{deltaiterationsMATLABcode}} \\
\hspace*{2.1cm} $\mbf{x}_* = \argmin \frac{1}{2}\|\mbf{A}\mbf{x} - \mbf{b}\|^2 + \lambda M_{h}^{\mu}(\mbf{x})$
}
\SetKwInOut{Input}{Input}
\SetKwInOut{Output}{Output}
\SetKwInOut{Return}{Return}
{\small\Input{$\mathbf{A}, \mbf{b}, \lambda, \mu$, $\textrm{prox}_{\mu h} (x)$

\Output{$\mbf{x}_*$}
     $\mathbf{K} = (\mbf{A}^T\mbf{A} + \frac{\lambda}{\mu}\mbf{I})^{-1}$ \tcp*{Regularized Inverse}

     $\mathbf{x}_\textrm{RLS} = \mathbf{K}\mbf{A}^T\mbf{b}$
     \tcp*{Regularized  Least Squares}

     $\boldsymbol{\delta}^{\textrm{new}} = \textrm{zeros(size(}\mathbf{x}_\textrm{RLS}))$ \tcp*{initialization}

    \Repeat{ $\|\boldsymbol{\delta}^{\textrm{new}} - \boldsymbol{\delta}^{\textrm{old}} \|_\infty < \epsilon$}
    {
      $\boldsymbol{\delta}^{\textrm{old}} = \boldsymbol{\delta}^{\textrm{new}} $ \tcp*{delta-iterations}

      $\boldsymbol{\delta}^{\textrm{new}} =
      \frac{\lambda}{\mu}
      \mathbf{K}\textrm{prox}_{\mu h}(\boldsymbol{\delta}^{\textrm{old}} + \mathbf{x}_\textrm{RLS})$;
    }

	\Return{$\mathbf{x}_* = \mathbf{x}_\textrm{RLS} + \boldsymbol{\delta}^{\textrm{new}}$}
}}	
\end{algorithm}

\textbf{Extension to Other Cost Functions:}
A scalar function is a Moreau envelope if it is convex, continuously differentiable and its derivative is Lipschitz continuous with a Lipschitz constant no greater than $1/\mu$. Here, $\mu>0$ is the smoothing parameter of the Moreau envelope.
Many cost functions, beyond Huber's and the $\epsilon$-insensitive cost functions, satisfy these conditions. To apply delta-iterations to other cost functions satisfying the Moreau envelope condition, we only need to identify the corresponding proximal mapping. The derivative relation for the Moreau envelope is key to this task \cite[Ch. 6.7.2]{BeckFOoptBook}:
\be
\frac{d}{dx}M_h^{\mu}(x) = \frac{1}{\mu}(x - \textrm{prox}_{\mu h}(x))
\label{ddxMenv}
\ee
To illustrate the retrieval of the proximal mapping, let
\be
J(x) =
\left\{
\begin{array}{lc}
0 & x \geq 0 \\
\frac{1}{2} x^2  & x \leq 0
\end{array}
\right.
\ee
be the one-sided cost function which penalizes only negative values of its argument. The function $J(x)$ is a valid Moreau envelope, since it is convex, continuously differentiable and has a second derivative bounded by $1$. Hence, the Moreau envelope condition is satisfied for any $0<\mu\leq 1$.

Given a Moreau envelope, the corresponding proximal mapping can be recovered from \mref{ddxMenv} as:
\be
\textrm{prox}_{\mu h}(x) = x - \mu\frac{d}{dx}J(x) =
\left\{
\begin{array}{lc}
x & x \geq 0 \\
(1-\mu) x  & x \leq 0
\end{array}
\right.
\label{prox-onesided}
\ee
If one-sided cost function is used in an application (see the numerical results section), we can choose any $\mu$ in $(0,1]$, say $\mu = 1/2$, in \mref{prox-onesided} and use the resulting proximal mapping in the delta-iteration solver. Note that, when the cost function is a valid Moreau envelope, delta-iterations can be applied without explicitly identifying the seed function $h(u)$ that generates the envelope.

\begin{algorithm}
\label{AlgMinMenvNorm}
\caption{\small{Min. Moreau Envelope Solution (see \cite{deltaiterationsMATLABcode})} \\
\hspace*{2.1cm} $\mbf{x}_* = \argmin
M_h^\mu(\mathbf{x})$ s.t. $\mathbf{A}\mathbf{x} = \mathbf{b}$
}
\SetKwInOut{Input}{Input}
\SetKwInOut{Output}{Output}
\SetKwInOut{Return}{Return}
{\small \Input{$\mathbf{A} \in \mathbb{R}^{N \times K} (K > N), \mbf{b}, \textrm{prox}_{\mu h} (x)$

\Output{$\mbf{x}_*$}
     $\mathbf{A}^{\dagger} = \mathbf{A}^T(\mathbf{A}\mathbf{A}^T)^{-1}$ \tcp*{Right pseudo-inverse}

     $\mathbf{x}_\textrm{MN} = \mathbf{A}^{\dagger}\mbf{b}$
     \tcp*{Minimum norm soln.}

     $\boldsymbol{\delta}^{\textrm{new}} = \textrm{zeros(size(}\mathbf{x}_\textrm{MN}))$ \tcp*{initialization}

    \Repeat{ $\|\boldsymbol{\delta}^{\textrm{new}} - \boldsymbol{\delta}^{\textrm{old}} \|_\infty < \epsilon$}
    {
      $\boldsymbol{\delta}^{\textrm{old}} = \boldsymbol{\delta}^{\textrm{new}}$ \tcp*{delta-iterations}

      $\mathbf{u} = \textrm{prox}_{\mu h}(\boldsymbol{\delta}^{\textrm{old}} + \mathbf{x}_\textrm{MN})$;

      $\boldsymbol{\delta}^{\textrm{new}} =
      \mathbf{u} - \mathbf{A}^{\dagger}(\mathbf{A}\mathbf{u})$; \tcp*{$\mathcal{O}(NK)$ update}
    }

	\Return{$\mathbf{x}_* = \mathbf{x}_\textrm{MN} + \boldsymbol{\delta}^{\textrm{new}}$}
}}	
\end{algorithm}


\begin{figure}[t]
\centering
\includegraphics[height=9cm]{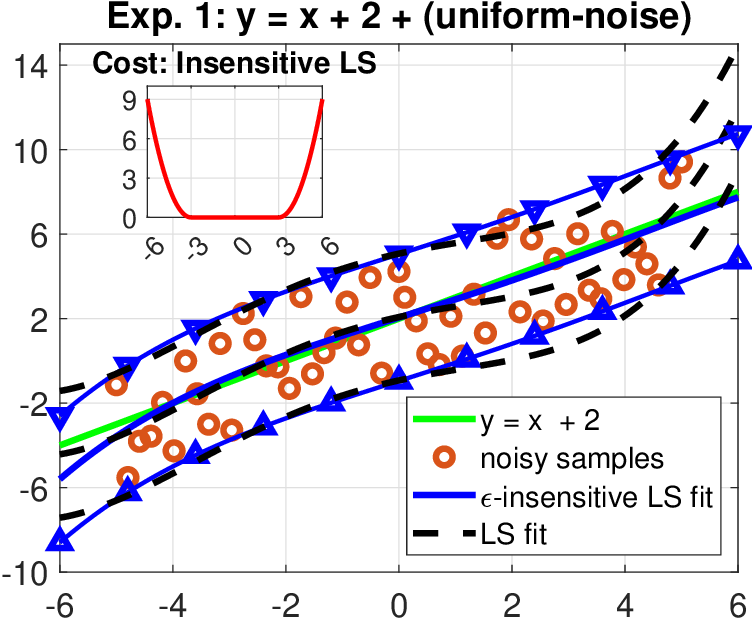}
\caption{The least squares (LS) and $\epsilon$-insensitive fit to the data. The samples in between triangle marked curves are located in the insensitive part of the cost function and do not contribute to the total cost.}
\label{figtubefit}
\end{figure}

\begin{figure}[t]
\centering
\includegraphics[height=9cm]{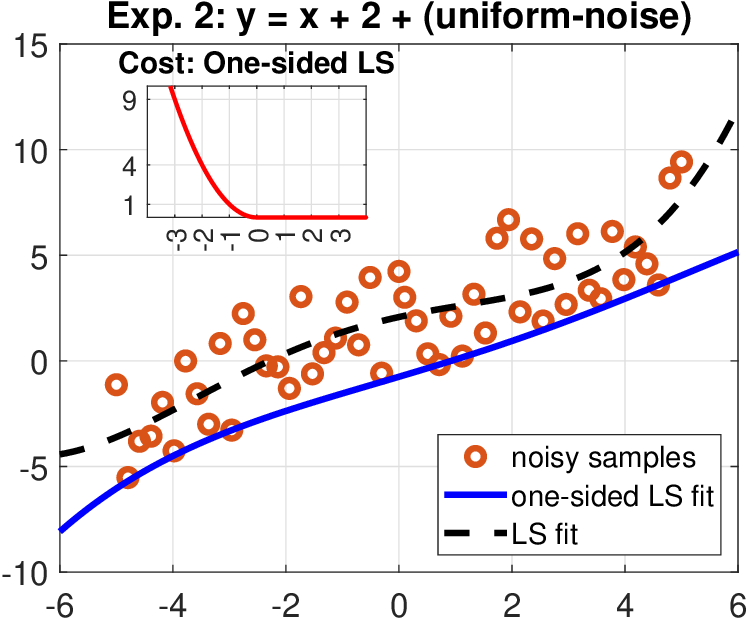}
\caption{One-sided least squares fit penalizing only the negative valued residual error. In the final fit, all samples are located above the regression function leading to a total cost of 0.}
\label{figsupfit}
\end{figure}

The goal of these three experiments is to estimate the parameters of fourth-order polynomial models using different cost functions. The last two experiments are more specialized and concern non-negative least-squares (NNLS) problems.

\vspace*{-0.4cm}
\section{Numerical Results}
We present five experiments to illustrate different use cases of the proposed method. We generate $N=50$ input samples, $x_i$, uniformly distributed over the interval $[-5,5]$, and obtain noisy observations according to $y = x + 2 + \textrm{(noise)}$. The goal of these three experiments is to estimate the parameters of fourth-order polynomial models using different cost functions. The last two experiments are more specialized and related to the non-negative least squares (NNLS) problems. The stopping criteria $\epsilon$ for delta-iterations is set as $\epsilon = 10^{-10}$ in all experiments.

\textbf{Experiment 1:} The noise is uniformly distributed over $[-3,3]$. We construct an insensitive least-squares (LS) fit by choosing $L_\lambda(x)$ from Table~\ref{TableM} as the cost function, with $\lambda=3$. Figure~\ref{figtubefit} shows the result obtained using delta-iterations (Algorithm~\ref{AlgSQUAREM}) and the conventional LS solution, which serves as the initial estimate for delta-iterations. The insensitive LS fit (blue curve) can be interpreted as fitting a tube, rather than a curve, to the data. The tube is centered on the fitted polynomial and its boundaries are indicated by the triangle markers in Figure~\ref{figtubefit}. Samples lying within the tube incur zero cost. Since the noise is bounded in this experiment, all observations can be accommodated within the tube when $\lambda=3$. Hence, the optimal cost is zero. We note that the SQUAREM-accelerated delta-iterations (Algorithm~\ref{AlgSQUAREM}) and the non-accelerated delta-iterations (Algorithm~\ref{Alg1}) converge in 8 and 856 iterations, respectively.

\textbf{Experiment 2:} We use the same noisy data realization of Experiment 1; but replace the insensitive LS cost with the one-sided LS cost ($J_\mu(x)$ in Table~\ref{TableM}). Applying Alg.~\ref{AlgSQUAREM} with the proximal mapping of $J_\mu(x)$ yields the result in Figure~\ref{figsupfit}. We observe that all samples lie above the fitted curve and have non-negative residuals. Hence, the minimized cost is exactly zero. We note that there are infinitely many solutions to this problem and SQUAREM (Alg.~\ref{AlgSQUAREM}) converged to one of them with all non-negative residuals in 7 iterations.

\begin{figure}[t]
\centering
\includegraphics[height=9cm]{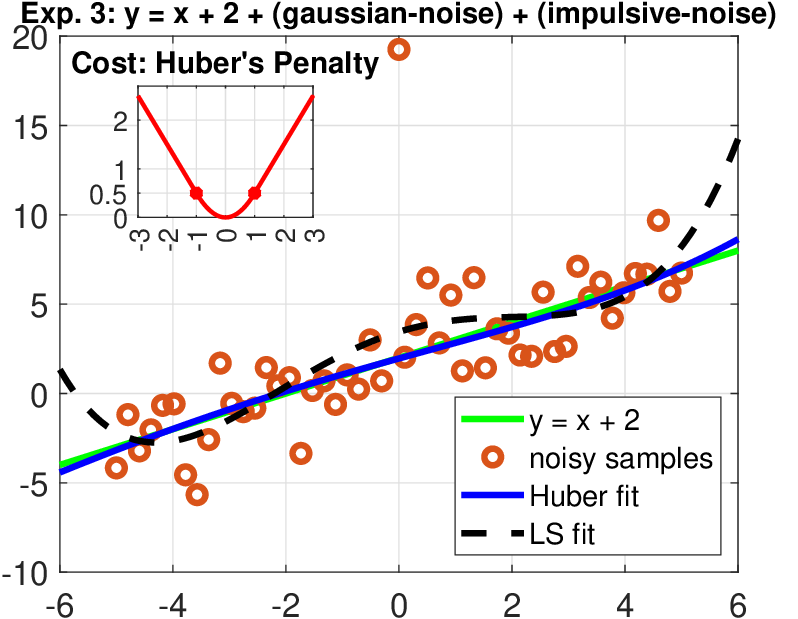}
\caption{Robust regression via Huber's loss. Data has an impulsive noise component at $x=0$.}
\label{fighuberfit}
\end{figure}

\textbf{Experiment 3:} We use the same problem setting as in the first two experiments, but have Gaussian plus impulsive noise. As shown in Figure~\ref{fighuberfit}, the impulsive noise affects a single sample located at $x=0$. We use Huber's penalty $H_\mu(x)$ with $\mu=1$ for robust regression. Running delta-iterations (Alg.~\ref{AlgSQUAREM}) with the corresponding proximal mapping yields the result in Figure~\ref{fighuberfit}. The SQUAREM-accelerated (Alg.~\ref{AlgSQUAREM}) and non-accelerated (Alg.~\ref{Alg1}) delta-iterations converged in 10 and 63 iterations, respectively.

\begin{figure}[t]
\centering
\includegraphics[height=10cm]{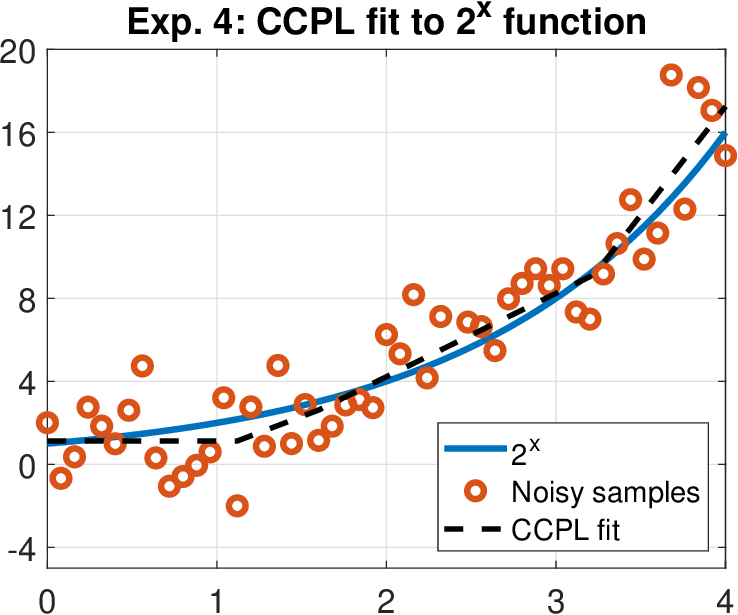}
\caption{Continuous Convex Piecewise-Linear (CCPL) fit by an approximate solution of a Non-Negative Least Squares (NNLS) problem with Algorithm~\ref{AlgMenvReg}.}
\label{figexpfit}
\end{figure}

\textbf{Experiment 4:} We study continuous convex piecewise-linear (CCPL) data fitting \cite[Sec. 6.5.5]{BoydBook}. Figure~\ref{figexpfit} shows 51 noisy samples of $2^x$ function, obtained by uniformly sampling $x$ over $[0,4]$ with the sampling period of $0.08$ units. The goal is to determine the best fit to the observations taking into account the {\em continuity and convexity}.

We construct a LASSO-like problem using ramp functions (also known as ReLU), $(x-c_l)_+ = \max(x-c_l,0)$, with corner locations $c_l$ serving as the dictionary atoms in an overcomplete representation \cite{LASSOBook}. The hypothesis class ${\cal{F}}$ consists of linear combinations of ramp functions and is therefore continuous and piecewise-linear by construction: ${\cal{F}} = \{ f(x) : f(x) = \alpha + \sum_{l=1}^{K} a_l \left(x-c_l\right)_+ \}$. To ensure convexity, we constrain the ramp coefficients to be nonnegative, i.e., $a_l \geq 0$. The coefficient $\alpha$, which represents the constant(bias) term, is left unconstrained.

We construct an underdetermined equation system $\mbf{A}\mbf{x} = \mbf{b}$ corresponding to this overcomplete representation as follows. Each column of $\mbf{A}$ represents a dictionary atom, namely, a ramp function with corner location $c_l$, evaluated at the 51 uniformly spaced points in $[0,4]$. We choose 201 corner locations uniformly spaced over $[0,4]$, with an interval of $0.02$, forming a $51 \times 201$ dictionary matrix.

We formulate the LASSO-like representation problem as follows
\be
\min_{\mathbf{x}} \frac{1}{2}\|\mbf{A}\mbf{x} - \mbf{b}\|^2 + \lambda J_{h}^{\mu}(\mbf{x})
\label{expfitproblem}
\ee
where the one-sided loss function $J_h^\mu(\mbf{x})$ (see Table~\ref{TableM}) penalizes negative ramp coefficients. Figure~\ref{figexpfit} shows the CCPL fit obtained using Algorithm~\ref{AlgMenvReg} with $\lambda=10^6$ and $\mu=1$.


\textbf{Experiment 5:} We study the minimum $\ell_1$-norm solution of an underdetermined linear system:
\be
\min \|\mbf{x}\|_1 \quad \textrm{s.t.} \quad
\left[
\begin{array}{cccc}
1 & 3 & 2 & 4 \\
2 & 3 & 2 & 3 \\
\end{array}
\right]
\mbf{x}
=
\left[
\begin{array}{c}
2 \\
3
\end{array}
\right].
\ee

The exact solution to this problem is $\mbf{x}_* =
\begin{bmatrix} 1 & 1/3 & 0 & 0 \end{bmatrix}^{T}$. A numerical solution can be obtained by linear programming or other specialized solvers \cite{cadzowl1norm}. Here, instead of minimizing the $\ell_1$ norm directly, we use Algorithm~\ref{AlgMinMenvNorm} with Huber's penalty $H_\mu(x)$ to obtain an approximate solution.

\begin{figure}[t]
\centering
\includegraphics[height=9cm]{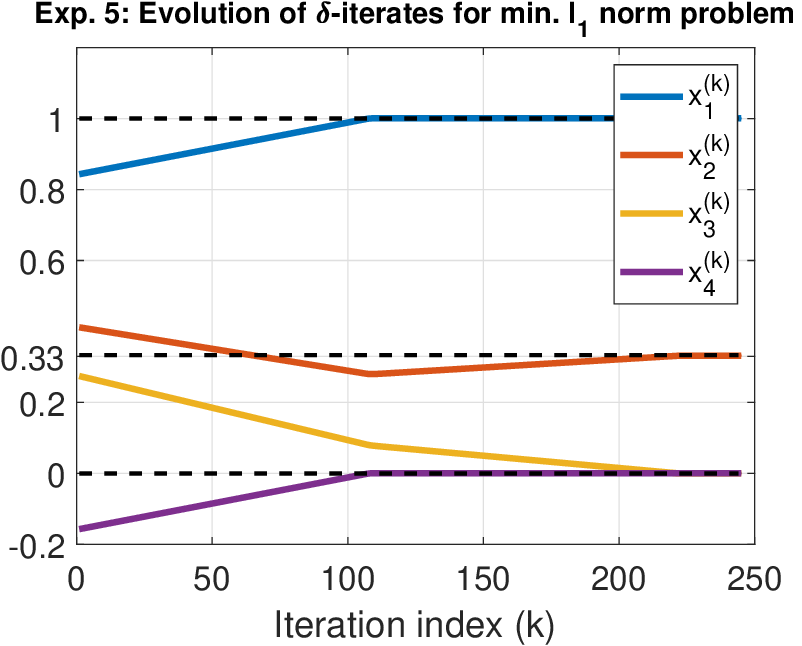}
\caption{Iterates for the approximate solution of $\ell_1$-norm minimization problem with Algorithm~\ref{AlgMinMenvNorm}.}
\label{figminorm}
\end{figure}

Figure~\ref{figminorm} shows the evolution of the delta-iterates for $\mu=10^{-3}$.  The dashed lines indicate the values of the exact solution. The figure illustrates how delta-iterations move from the conventional minimum $\ell_2$-norm solution toward the minimum $\ell_1$-norm solution.

For an underdetermined system with an $N\times K$ matrix $\mbf{A}$, where $K>N$, there exists a minimum $\ell_1$-norm solution with at most $N$ nonzero coefficients \cite[Sec. 3.1.1]{DonohoAtomic2001}. Hence, at least $K-N$ coefficients are zero. Once the zero-valued coefficients have been identified, the corresponding columns can be eliminated from $\mbf{A}$, leaving a system with $N$ equations and $N$ unknowns which can be solved for the nonzero coefficients. Thus, the main challenge is to identify the active (nonzero) coefficients among the $K$ candidates.  Figure~\ref{figminorm} suggests that delta-iterations can be terminated early as soon as the zero-valued coefficients can be reliably detected, say around the 120th iteration in this example.

\vspace*{-0.4cm}
\section{Summary and Conclusions}
From a practical viewpoint, the proposed delta-iterations start from the conventional LS solution and determine the perturbation vector $\boldsymbol{\delta}$ required to reach the solution of a different optimization problem, such as robust regression or non-negative least squares (NNLS). Their convergence guarantee, simplicity of implementation, and generality make delta-iterations an attractive practical tool. Ready-to-run MATLAB code for reproducing all the numerical experiments is available at \cite{deltaiterationsMATLABcode}.

From a theoretical viewpoint, the main technical contribution of the proposed method is to incorporate the variational definition of the Moreau envelope into the original optimization problem. This reformulation introduces an auxiliary variable $\mbf{u}$ and allows us to combine the two quadratic terms (the least-squares term and the quadratic term in the Moreau envelope definition) before minimizing with respect to $\mbf{u}$. A similar tractable approach may also be applicable to related problems involving Bregman distances instead of quadratic distances, as discussed in \cite[Ch. 9.1]{BeckFOoptBook}.

\vspace*{-0.4cm}
{\small

}

\appendix
%
%
%

\section{$\ell_2$ regularized Linear Regression with Minimum Moreau Envelope (Alg.~\ref{AlgSQUAREM})}
This section gives the derivation for the delta-iteration solution of
\begin{equation}
    J(\mathbf{x}) = M_h^\mu(\mathbf{b} - \mathbf{Ax}) + \lambda \|\mathbf{x}\|^2
\label{appMr1}
\end{equation}
where $\mu > 0$ is the Moreau envelope parameter and $\lambda > 0$ is the regularization parameter of the $\ell_2$ norm regularizer.

\textbf{Step 1: Joint Optimization Formulation}
The Moreau envelope of $h(\mathbf{u})$ evaluated at $\mathbf{b} - \mathbf{Ax}$ is written as:
\begin{equation}
    M_h^\mu(\mathbf{b} - \mathbf{Ax}) = \min_u \left( h(\mathbf{u}) + \frac{1}{2\mu}\|(\mathbf{b - Ax}) - \mathbf{u}\|^2 \right).
\end{equation}
Substituting this variational relation into \mref{appMr1}, the objective function of the joint optimization problem over $\mathbf{x}$ and $\mathbf{u}$ becomes:
\begin{equation}
    J(\mathbf{x}, \mathbf{u}) = h(\mathbf{u}) + \frac{1}{2\mu} \|\mathbf{Ax} - (\mathbf{b} - \mathbf{u})\|^2 + \lambda \|\mathbf{x}\|^2.
\label{appMr2}
\end{equation}

\textbf{Step 2: Optimizing over $\mathbf{x}$}
To optimize over $\mathbf{x}$ for a fixed $\mathbf{u}$, we calculate the gradient of $J(\mathbf{x}, \mathbf{u})$  wrt $\mathbf{x}$ and equate the result to $\mathbf{0}$:
\begin{align}
    \frac{1}{\mu} \mathbf{A}^T (\mathbf{Ax - (b - u)}) + 2\lambda \mathbf{x} &= \mathbf{0}.
\label{appMroptc}
\end{align}
The optimal $\mathbf{x}$ for a given $\mathbf{u}$ is then
\begin{equation}
    \mathbf{x}_*(\mathbf{u}) = (\mathbf{A}^T \mathbf{A} + 2\mu\lambda \mathbf{I})^{-1} \mathbf{A}^T \mathbf{(b - u)}
\end{equation}
We denote regularized pseudo-inverse matrix as $\mathbf{K} = (\mathbf{A}^T\mathbf{A} + 2\mu\lambda \mathbf{I})^{-1} \mathbf{A}^T$. Then $\mathbf{x}_*(\mathbf{u})$ becomes
\begin{equation}
    \mathbf{x}_*(u) = \mathbf{Kb - Ku} = \mathbf{x}_{RLS} - \boldsymbol{\boldsymbol{\delta}}
\end{equation}
where $\mathbf{x}_{RLS} = \mathbf{Kb}$ and $\boldsymbol{\delta} = \mathbf{Ku}$.

\textbf{Step 3: Proximal Gradient Iterations for $\mathbf{u}$}
Substituting $\mathbf{x}_*(u)$ back into \mref{appMr2}, we
get
\be
J(\mathbf{x}_*,\mathbf{u}) = h(\mathbf{u}) + \frac{1}{2\mu} \|\mathbf{Ax_*} - (\mathbf{b} - \mathbf{u})\|^2 + \lambda \mathbf{x}^T_* \mathbf{x_*}.
\ee
Without expanding the square, inserting $\lambda\mathbf{x}_* = -\frac{1}{2\mu}(\mathbf{A}^T(\mathbf{A\mathbf{x}_*-(b-u)}))$ which is derived from the optimality condition \mref{appMroptc}, we get
\be
J(\mathbf{x}_*,\mathbf{u}) = h(\mathbf{u}) + \frac{1}{2\mu} \|\mathbf{Ax_*} - (\mathbf{b} - \mathbf{u})\|^2
-\frac{1}{2\mu}(\mathbf{Ax_* -(b-u)})^T\mathbf{A}\mathbf{x}_*.
\label{appMr3}
\ee
Denoting $\mathbf{y} = \mathbf{(b-u)}$, we add and subtract $\mathbf{y}$ from the last term on the right side of \mref{appMr3} to get
\be
J(\mathbf{x}_*,\mathbf{u}) & =  & h(\mathbf{u}) + \frac{1}{2\mu} \|\mathbf{Ax_*} - \mathbf{y}\|^2
-\frac{1}{2\mu}(\mathbf{Ax_* -\mathbf{y}})^T(\mathbf{A}\mathbf{x}_* - \mathbf{y} + \mathbf{y}) \nn \\
& = &  h(\mathbf{u}) -
\frac{1}{2\mu}(\mathbf{Ax_*} -\mathbf{y})^T \mathbf{y}
\nn \\
& = & h(\mathbf{u}) -
\frac{1}{2\mu}(\mathbf{AKy} -\mathbf{y})^T \mathbf{y}
\nn \\
& = & h(\mathbf{u})
+ \frac{1}{2\mu}\mathbf{y}^T(\mathbf{I - AK})\mathbf{y}.
\label{appMr4}
\ee
Denoting the smooth part of objective function in \mref{appMr4} as
\begin{equation}
    g(\mathbf{u}) = \frac{1}{2\mu} \mathbf{(b - u)}^T (\mathbf{I - AK}) (\mathbf{b - u})
\end{equation}
and calculating its gradient, we get:
\begin{equation}
    \nabla g(\mathbf{u}) =
    \frac{1}{\mu} \mathbf{(I - AK)} (\mathbf{u} - \mathbf{b})
\end{equation}
Similar to the gradient descent, in the proximal gradient descent  framework, the monotonic decrease of the cost function is guaranteed if the step-size bounded by the 2 times the reciprocal Lipschitz constant $L$ of the gradient function (or the maximum eigenvalue of the Hessian matrix of $g(\mbf{u})$). Hence, selecting the step size as $\alpha = \mu$, the descent update at the k'th iteration $\mathbf{z}^{(k)}$ becomes:
\begin{align}
    \mathbf{z}^{(k)} &= \mathbf{u}^{(k)} - \mu \nabla g(\mathbf{u}^{(k)}) \nonumber \\
            &= \mathbf{u}^{(k)} - \mathbf{(I - AK)}(\mathbf{u}^{(k)} - \mathbf{b}) \nonumber \\
            &= \mathbf{A}(\mathbf{K} \mathbf{u}^{(k)}) + \mathbf{(b - AK b)}
\end{align}
Substituting $\boldsymbol{\delta}^{(k)} = \mathbf{K} \mathbf{u}^{(k)}$ and the residual of regularized least squares $\mathbf{r}_{RLS} = \mathbf{b - A x}_{RLS} = \mathbf{b} - \mathbf{AKb}$, we get
\begin{equation}
    \mathbf{z}^{(k)} = \mathbf{A}\boldsymbol{\delta}^{(k)} + \mathbf{r}_{RLS}.
\end{equation}

\textbf{Step 4: Proximal Update with $\boldsymbol{\delta}$ Iterations}
In the proximal update step, $\mathbf{z}^{(k)}$ is mapped with the proximal operator of the non-smooth function  $h(\mbf{u})$ to get the next iterate:
\begin{equation}
    \mathbf{u}^{(k+1)} = \text{prox}_{\mu h}(\mathbf{z}^{(k)}) = \text{prox}_{\mu h}(\mathbf{A}\boldsymbol{\delta}^{(k)} + \mathbf{r}_{RLS})
\end{equation}
Applying $\mbf{K}$ from left, we finalize the delta-iterations:
\begin{equation}
    \boldsymbol{\delta}^{(k+1)} = \mathbf{K} \text{prox}_{\mu h}(\mathbf{A}\boldsymbol{\delta}^{(k)} + \mathbf{r}_{RLS}).
\end{equation}
The SQUAREM-accelerated version of this algorithm is given in Algorithm~\ref{AlgSQUAREM}.


\section{Least Squares Linear Regression with Moreau Envelope Regularization (Alg.~\ref{AlgMenvReg})}
We consider the conventional linear regression problem regularized by the Moreau envelope of $h(\mathbf{u})$
\begin{equation}
    \min_x \frac{1}{2} \|\mathbf{Ax - b}\|^2 + \lambda M_h^\mu(\mathbf{x})
\end{equation}
where $M_h^\mu(x)$ is the Moreau envelope of $h(\mathbf{u})$ with smoothing parameter $\mu > 0$.

\textbf{Step 1: Joint Optimization Formulation}
Substituting the definition of the Moreau envelope, we get a joint optimization problem over variables $\mathbf{x}$ and $\mathbf{u}$ whose objective function is:
\begin{equation}
    J(\mathbf{x}, \mathbf{u}) = \frac{1}{2} \|\mathbf{Ax - b}\|^2 + \frac{\lambda}{2\mu} \|\mathbf{x - u}\|^2 + \lambda h(\mathbf{u}).
\label{appCLSMenvR1}
\end{equation}

\textbf{Step 2: Optimizing over $\mathbf{x}$}
To optimize over $\mathbf{x}$ for a fixed $\mathbf{u}$, we calculate the gradient of $J(\mathbf{x}, \mathbf{u})$  wrt $\mathbf{x}$ and equate the result to $\mathbf{0}$:
\begin{align}
    \nabla_\mathbf{x} J(\mathbf{x}, \mathbf{u}) = \mathbf{A}^T(\mathbf{Ax - b}) + \frac{\lambda}{\mu}(\mathbf{x} - \mathbf{u}) = \mbf{0}
\label{appCLSMenvR2}
\end{align}
The optimal $\mathbf{x}$ for a given $\mathbf{u}$ is then
\begin{equation}
    \mathbf{x}_*(\mathbf{u}) = \left( \mathbf{\mbf{A}^T A} + \frac{\lambda}{\mu} \mathbf{I} \right)^{-1} \left( \mathbf{A}^T \mathbf{b} + \frac{\lambda}{\mu} \mathbf{u} \right)
\label{appCLSMenvR3}
\end{equation}

\textbf{Step 3: Proximal Gradient Iterations for $\mathbf{u}$}
Substituting $\mathbf{x}_*(u)$ back into \mref{appCLSMenvR1}, we get
\be
J(\mathbf{x}_*,\mathbf{u}) = \lambda h(\mathbf{u})
+
\frac{1}{2} \|\mathbf{A\mathbf{x}_* - b}\|^2 + \frac{\lambda}{2\mu} \|\mathbf{\mathbf{x}_* - u}\|^2.
\ee
Without expanding the square, inserting $\mathbf{A}^T(\mathbf{Ax_* - b})  = - \frac{\lambda}{\mu}(\mathbf{x}_* - \mathbf{u})
$, which is derived from the optimality condition \mref{appCLSMenvR2}, we get
\be
J(\mathbf{x}_*,\mathbf{u}) & = &   \lambda h(\mathbf{u})
+
\frac{1}{2} (\mathbf{A\mathbf{x}_* - b})^T
(\mathbf{A\mathbf{x}_* - b})
+
\frac{\lambda}{2\mu} \|\mathbf{\mathbf{x}_* - u}\|^2. \nn \\
& = & \lambda h(\mathbf{u}) +
\frac{1}{2}\mathbf{x}_*^T \left(-\frac{\lambda}{\mu}(\mathbf{x}_* - \mathbf{u})\right)
-\frac{1}{2}\mathbf{b}^T(\mathbf{A\mathbf{x}_* - \mathbf{b}})
+
\frac{\lambda}{2\mu} \|\mathbf{\mathbf{x}_* - u}\|^2 \nn \\
& = & \lambda h(\mathbf{u}) +
\frac{\lambda}{2\mu}(-\mathbf{u})^T (\mathbf{x}_* - \mathbf{u})
-\frac{1}{2}\mathbf{b}^T\mathbf{A\mathbf{x}_*} + \frac{1}{2} \|\mathbf{b}\|^2 \nn \\
& = & \lambda h(\mathbf{u})
-\frac{1}{2}
\left(\frac{\lambda}{\mu}\mbf{u}
+\mathbf{A}^T\mathbf{b}\right)^T\mathbf{x}_*
+ \frac{\lambda}{2\mu} \|\mathbf{u}\|^2
+ \frac{1}{2} \|\mathbf{b}\|^2
\ee
Defining $\mathbf{M} = \mathbf{ A^T A} + \frac{\lambda}{\mu} \mathbf{I}$ and inserting $\mathbf{x}_* = \mathbf{M}^{-1}(\mathbf{A}^T\mathbf{b} +
\frac{\lambda}{\mu}\mbf{u})$ (see \mref{appCLSMenvR3})
\be
J(\mathbf{x}_*,\mathbf{u})
& = & \lambda h(\mathbf{u})
-\frac{1}{2}
(\mathbf{A}^T\mathbf{b} +
\frac{\lambda}{\mu}\mbf{u})^T
\mbf{M}^{-1}
(\mathbf{A}^T\mathbf{b} +
\frac{\lambda}{\mu}\mbf{u})
+ \frac{\lambda}{2\mu} \|\mathbf{u}\|^2
+ \frac{1}{2} \|\mathbf{b}\|^2 \nn \\
& = & \lambda h(\mathbf{u})
+ \frac{\lambda}{2\mu}\mbf{u}^T
\left(-\frac{\lambda}{\mu}\mathbf{M}^{-1} + \mbf{I}\right)\mbf{u}
-\frac{\lambda}{\mu}
(\mathbf{A}^T\mathbf{b})^T
\mbf{M}^{-1}\mbf{u}
+ c
\ee
where $c$ in the last line refers to the constants not affecting subsequent steps. Expressing the identity matrix appearing in the middle term as $\mbf{I} = \mbf{M}^{-1}\mbf{M}$, we get
\be
J(\mathbf{x}_*,\mathbf{u})
& = & \lambda h(\mathbf{u})
+ \frac{\lambda}{2\mu}\mbf{u}^T
\left(-\frac{\lambda}{\mu}\mathbf{M}^{-1} + \mbf{M}^{-1}\mbf{M}\right)\mbf{u}
-\frac{\lambda}{\mu}
(\mathbf{A}^T\mathbf{b})^T
\mbf{M}^{-1}\mbf{u}
+ c  \nn \\
& = & \lambda h(\mathbf{u})
+ \frac{\lambda}{2\mu}\mbf{u}^T \mbf{M}^{-1}
\left(-\frac{\lambda}{\mu}\mathbf{I} + \mbf{M}\right)\mbf{u}
-\frac{\lambda}{\mu}
(\mathbf{A}^T\mathbf{b})^T
\mbf{M}^{-1}\mbf{u}
+ c  \nn \\
& = & \lambda h(\mathbf{u})
+ \frac{\lambda}{2\mu}\mbf{u}^T \mbf{M}^{-1}
(\mathbf{A}^T \mathbf{A})
\mbf{u}
-\frac{\lambda}{\mu}
\mathbf{b}^T \mathbf{A}
\mbf{M}^{-1}\mbf{u}
+ c  \nn
\ee
Denoting the smooth part (quadratic part) of the  $J(\mathbf{x}_*,\mathbf{u})$ as $g(\mathbf{u})$ and calculating its gradient, we get:
\begin{equation}
    \nabla g(u) = \frac{\lambda}{\mu} \mathbf{M}^{-1} \mathbf{A}^T \mathbf{A u} - \frac{\lambda}{\mu} \mathbf{M}^{-1} \mathbf{A}^T \mathbf{b}
\end{equation}
To simplify this, we express $\mathbf{A}^T \mathbf{A}$ as $\mathbf{M} - \frac{\lambda}{\mu} \mathbf{I}$ and rewrite the gradient as
\be
    \nabla g(u) = \frac{\lambda}{\mu} (\mbf{I}-\frac{\lambda}{\mu}\mathbf{M}^{-1}) \mathbf{u} - \frac{\lambda}{\mu} \mathbf{M}^{-1} \mathbf{A}^T \mathbf{b}
\ee
We apply the proximal gradient method with the simplified gradient.  The maximum eigenvalue of $\nabla^2 g(u)$ is bounded by $\frac{\lambda}{\mu}$, allowing for a step size $\alpha = \frac{\mu}{\lambda}$. The gradient update  $\mathbf{z}^{(k)} = \mathbf{u}^{(k)} - \alpha \nabla g(\mathbf{u}^{(k)})$ then becomes:
\begin{align*}
    \mathbf{z}^{(k)} &= \mathbf{u}^{(k)} - \frac{\mu}{\lambda} \left[ \frac{\lambda}{\mu} \left( \mathbf{I} - \frac{\lambda}{\mu} \mbf{M}^{-1} \right) \mathbf{u}^{(k)} - \frac{\lambda}{\mu} \mbf{M}^{-1} \mathbf{A}^T \mathbf{b} \right] \\
     &= \mathbf{u}^{(k)} - \mathbf{u}^{(k)} + \frac{\lambda}{\mu} \mbf{M}^{-1} \mathbf{u}^{(k)} + \mbf{M}^{-1} \mathbf{A}^T \mathbf{b} \\
     & = \mbf{M}^{-1}
     \left( \mathbf{A}^T \mathbf{b} +
     \frac{\lambda}{\mu} \mathbf{u}^{(k)} \right)
\end{align*}

In the final step, the proximal operator of the non-smooth function scaled by $\alpha$ (step-size) is utilized. Since the step-size is $\alpha = \frac{\mu}{\lambda}$ and the non-smooth term is $\lambda h(\mbf{u})$, we have $\alpha \lambda h(\mbf{u}) = \mu h(\mbf{u})$ leading to:
\be
 \mbf{u}^{(k+1)} = \text{prox}_{\mu h} \left( \mathbf{z}^{(k)} \right)
 =
 \text{prox}_{\mu h} \left( \mbf{M}^{-1}
     \left( \mathbf{A}^T \mathbf{b} +
     \frac{\lambda}{\mu} \mathbf{u}^{(k)} \right)  \right).
\label{appCLSMenvR4}
\ee

\textbf{Step 4: Proximal Update with $\boldsymbol{\delta}$-Iterations}
We introduce $\mbf{x}_{RLS} = \mathbf{M}^{-1} \mathbf{A}^T \mathbf{b}$ and $\boldsymbol{\delta} = \frac{\lambda}{\mu}\mathbf{M}^{-1}\mbf{u}$ and express the optimal solution from \mref{appCLSMenvR3} as
\be
\mathbf{x}_*(\mathbf{u}_*)
& = &  \mbf{M}^{-1}\left( \mathbf{A}^T \mathbf{b} + \frac{\lambda}{\mu} \mathbf{u}_* \right)
\nn \\
& = &
\mbf{x}_{RLS} + \boldsymbol{\delta}_*
\ee
where $\mbf{u}_*$ is the fixed point of proximal mapping in \mref{appCLSMenvR4} and $\boldsymbol{\delta}_* = \frac{\lambda}{\mu}\mathbf{M}^{-1}\mbf{u}_*$ is associated parameter-space variable.

Left multiplying the fixed point mapping in \mref{appCLSMenvR4} with $\frac{\lambda}{\mu}\mathbf{M}^{-1}$, we reach the delta-iteration recursion:
\be
 \boldsymbol{\delta}^{(k+1)} =
 \text{prox}_{\mu h}
 (\mbf{x}_{RLS} + \boldsymbol{\delta}^{(k)}).
\ee
The algorithm is summarized in Algorithm~\ref{AlgMenvReg}.

\section{Minimum Moreau Envelope Solution of $\mathbf{Ax}=\mathbf{b}$ (Alg.~\ref{AlgMinMenvNorm})}
We consider the following problem
\begin{equation}
    \min_{\mathbf{x} \in \mathbb{R}^K} M_h^\mu(\mathbf{x}) \quad \text{subject to} \quad \mathbf{A}\mathbf{x} = \mathbf{b}
    \label{appAxeqb1}
\end{equation}
which reduces to the conventional minimum norm problem when $M_h^\mu(\mathbf{x})$ is replaced with $\ell_2$ norm square, that is $\|\mathbf{x}\|^2$. Here, $\mathbf{A}\mathbf{x} = \mathbf{b}$ is an under-determined equation system with fewer equations than unknowns, that is $\mbf{A}$ is $N \times K$ dimensional with $K>N$.

\textbf{Step 1: Joint Optimization Formulation}
Substituting the definition of the Moreau envelope, we get the following joint optimization problem over variables $\mathbf{x}$ and $\mathbf{u}$:
\begin{equation}
    \min_{\mathbf{x}, \mathbf{u} \in \mathbb{R}^K} \left\{ h(\mathbf{u}) + \frac{1}{2\mu} \|\mathbf{x} - \mathbf{u}\|_2^2 \right\} \quad \text{subject to} \quad \mathbf{A}\mathbf{x} = \mathbf{b}.
    \label{appAxeqb2pre}
\end{equation}

\textbf{Step 2: Optimizing over $\mathbf{x}$}
We optimize over $\mathbf{x}$ for a fixed $\mathbf{u}$. From \mref{appAxeqb2pre}, the optimization problem over $\mbf{x}$ for a fixed $\mbf{u}$ becomes
\begin{equation}
    \mathbf{x}^*(\mathbf{u}) = \argmin_{\mathbf{x} \in \mathbb{R}^K} \frac{1}{2\mu} \|\mathbf{x} - \mathbf{u}\|_2^2 \quad \text{subject to} \quad \mathbf{A}\mathbf{x} = \mathbf{b}.
    \label{appAxeqb2}
\end{equation}
The problem in \mref{appAxeqb2} is almost identical to the conventional minimum norm problem except the objective function is the norm square of the difference vector between $\mbf{x}$ and $\mbf{u}$ (the square of the distance between $\mbf{x}$ and $\mbf{u}$). To convert the problem to the usual minimum norm problem, we introduce $\mathbf{e} = \mathbf{x} - \mathbf{u}$ variable in \mref{appAxeqb2}
\be
    \mathbf{x}^*(\mathbf{u})
    & = &
    \argmin_{\mathbf{e} \in \mathbb{R}^K} \frac{1}{2\mu} \|\mathbf{e} \|_2^2 \quad \text{subject to} \quad \mathbf{A}\mathbf{e} = \mathbf{b} - \mathbf{A}\mathbf{u}
    \label{appAxeqb3}
\ee
The optimizer of \mref{appAxeqb3} is then $\mathbf{e}_* = \mathbf{A}^\dagger(\mathbf{b} - \mathbf{A}\mathbf{u})$ where $\mathbf{A}^\dagger = \mathbf{A}^T(\mathbf{A}\mathbf{A}^T)^{-1} \in \mathbb{R}^{K \times N}$ is the right-pseudoinverse matrix. The solution for $\mbf{x}$, then, becomes
$ \mathbf{x}_*(\mathbf{u})  = \mathbf{u} + \mathbf{e}_*$:
\be
    \mathbf{x}_*(\mathbf{u}) & = &  \mathbf{u} + \mathbf{A}^T(\mathbf{A}\mathbf{A}^T)^{-1}(\mathbf{b} - \mathbf{A}\mathbf{u})
    = \mathbf{u} + \mathbf{A}^\dagger(\mathbf{b} - \mathbf{A}\mathbf{u})
    \label{appAxeqb4}
\ee
\textbf{Step 3: Proximal Gradient Iterations for $\mathbf{u}$}
Substituting $\mathbf{x}_*(u)$ into \mref{appAxeqb2}, the optimization problem over $\mbf{u}$ becomes
\be
    \min_{\mathbf{u} \in \mathbb{R}^K}
    \left\{ h(\mathbf{u}) +
    \frac{1}{2\mu} \|\mathbf{x}_*(\mathbf{u}) - \mathbf{u}\|_2^2 \right\}.
    \label{appAxeqb5}
\ee
Expressing $\mathbf{x}_*(\mathbf{u}) - \mathbf{u}$ as
\be
\mathbf{x}_*(\mathbf{u}) - \mathbf{u}
& = &
\mathbf{A}^\dagger\mathbf{b} - \mathbf{A}^\dagger\mathbf{A}\mathbf{u}
\nn \\
& = &
\mathbf{x}_{\text{MN}} - \mathbf{P}_R \mathbf{u}
\ee
where $\mathbf{x}_{\text{MN}} = \mathbf{A}^\dagger\mathbf{b}$ is the conventional minimum norm solution and $\mathbf{P}_R =  \mathbf{A}^\dagger\mathbf{A} = \mathbf{A}^T(\mathbf{A}\mathbf{A}^T)^{-1} \mathbf{A} \in \mathbb{R}^{K \times K}$ is the projection matrix to the range space of $\mbf{A}^T$ (row space of $\mbf{A}$).

Since $\mathbf{x}_{\text{MN}}$ is in the row space of $\mbf{A}$ (or $\mathbf{P}_R \mathbf{A}^\dagger = \mathbf{A}^\dagger$), we have $\mathbf{P}_R \mathbf{x}_{\text{MN}} = \mathbf{x}_{\text{MN}}$  leading to
\be
\mathbf{x}_*(\mathbf{u}) - \mathbf{u}
& = &
\mathbf{P}_R (\mathbf{x}_{\text{MN}} -  \mathbf{u}).
\label{appAxeqb6}
\ee

Inserting \mref{appAxeqb6} into \mref{appAxeqb5}, we express the optimization problem over $\mbf{u}$ as
\be
    \min_{\mathbf{u} \in \mathbb{R}^K}
     \{ h(\mathbf{u}) + \underbrace{\frac{1}{2\mu} (\mathbf{u} - \mathbf{x}_{\text{MN}})^T \mathbf{P}_R (\mathbf{u} - \mathbf{x}_{\text{MN}})}_{g(\mathbf{u})}
    \}.
    \label{appAxeqb7}
\ee
Next, we apply the proximal gradient method for the solution of the problem $\min_{\mathbf{u}} \{h(\mbf{u}) + g(\mbf{u})\}$ with smooth $g(\mbf{u})$ and non-smooth $h(\mbf{u})$ functions.

The gradient of $g(\mbf{u})$ is
\begin{equation}
        \nabla g(\mathbf{u}) = \frac{1}{\mu} \mathbf{P}_R (\mathbf{u} - \mathbf{x}_{\text{MN}}) = \frac{1}{\mu} (\mathbf{P}_R \mathbf{u} - \mathbf{x}_{\text{MN}}).
\end{equation}

The Hessian of $g(\mathbf{u})$ is $\nabla^2 g(\mathbf{u}) = \frac{1}{\mu} \mathbf{P}_R$. Because $\mathbf{P}_R$ is an orthogonal projection matrix, its eigenvalues are either 0 or 1. Hence, the eigenvalues of the Hessian is bounded by $\frac{1}{\mu}$. Hence, we can select the step size $\alpha$ as $\alpha = \mu$. Following conventional gradient descent methodology, we express the gradient descent update as
\begin{align}
        \mathbf{z}^{(k)} &= \mathbf{u}^{(k)} - \mu \nabla g(\mathbf{u}^{(k)}) \nonumber \\
        &= \mathbf{u}^{(k)} - (\mathbf{P}_R \mathbf{u}^{(k)} - \mathbf{x}_{\text{MN}}) \nonumber \\
        &= \mathbf{x}_{\text{MN}} + (\mathbf{I}_K - \mathbf{P}_R)\mathbf{u}^{(k)}  \nn \\
        &= \mathbf{x}_{\text{MN}} + \mathbf{P}_R^\perp \mathbf{u}^{(k)}
        \label{appAxeqb8}
\end{align}
where $\mathbf{P}_R^\perp = \mathbf{I} - \mathbf{P}_R \in \mathbb{R}^{K \times K}$ is the projector to the orthogonal complement.

In the final step, the gradient update is mapped with proximal function to get $\mbf{u}$-recursion:
\begin{equation}
        \mathbf{u}^{(k+1)} = \textrm{prox}_{\mu h}(\mathbf{z}^{(k)}) = \textrm{prox}_{\mu h}\!\left(\mathbf{x}_{\text{MN}} + \mathbf{P}_R^\perp \mathbf{u}^{(k)}\right).
        \label{appAxeqb9}
\end{equation}
To develop delta-iterations, we re-write $\mbf{x}(u)$ relation given in \mref{appAxeqb4} as follows:
\be
    \mathbf{x}_*(\mathbf{u}) & = &  \mathbf{u} + \mathbf{A}^\dagger(\mathbf{b} - \mathbf{A}\mathbf{u}) \nn \\
    & = &  \mathbf{x}_{\text{MN}} + \mathbf{P}_R^\perp \mathbf{u}.
    \label{appAxeqb10}
\ee
Defining $\boldsymbol{\delta}^{(k)} = \mathbf{P}_R^\perp \mathbf{u}^{(k)}$ as the perturbation vector from the minimum norm solution, $\mbf{x}(u)$ becomes $\mbf{x}_*(\mathbf{u}^{(k)}) = \mathbf{x}_{\text{MN}} + \boldsymbol{\delta}^{(k)}$. Left multiplying \mref{appAxeqb9} with $\mathbf{P}_R^\perp$, we get delta-iterations:
\be
\boldsymbol{\delta}^{(k+1)}
= \mathbf{P}_R^\perp
\textrm{prox}_{\mu h}\!\left(\mathbf{x}_{\text{MN}} + \boldsymbol{\delta}^{(k)}\right)
\label{appAxeqb11}
\ee
Here $\boldsymbol{\delta}^{(k)}$ and $\mathbf{P}_R^\perp$ are $K \times 1$ and $K \times K$ dimensional objects, respectively. Since this problem involves an underdetermined equation system, that is $\mbf{A}$ of dimensions $N \times K$ with $K>N$,   the vectors and matrices in the $\mbf{u}$-iteration (see \mref{appAxeqb9}) and delta-iteration (as in \mref{appAxeqb11}) are in terms of larger dimension of the matrix $\mbf{A}$. To
reduce the computation from $\mathcal{O}(N^2)$ operations to $\mathcal{O}(NK)$ per iteration, we compute the delta-iterations in two steps
\begin{itemize}
\item[1.] $\mbf{t} = \textrm{prox}_{\mu h}\!\left(\mathbf{x}_{\text{MN}}
     + \boldsymbol{\delta}^{(k)}\right)$

\item[2.] $\boldsymbol{\delta}^{(k+1)} =
\mbf{t} - \mathbf{A}^\dagger(\mathbf{A}\mbf{t})$
\end{itemize}
The 2nd step involves two matrix multiplications each of which is $\mathcal{O}(NK)$ operations.

The algorithm is summarized in Algorithm~\ref{AlgMinMenvNorm}.

\end{document}